\documentclass[aps,12pt]{revtex4}
\usepackage[dvips]{epsfig}

\parskip=\medskipamount

\newcounter{fig}

\begin{document}

\title{Two-dimensional Site-Bond Percolation as an Example
of Self-Averaging System.}

\author{ Oleg A.
Vasilyev$^{[1,*]}$ }

\affiliation{L.D.Landau Institute for Theoretical Physics RAS,
117940 Moscow, Russia \\  e-mail vasilyev@itp.ac.ru}

\date{\today}

\bigskip

\begin{abstract}
The Harris-Aharony criterion for a statistical model predicts, that if 
a specific
heat exponent $\alpha \ge 0$, then this model does not exhibit
self-averaging. In two-dimensional percolation 
model the index $\alpha=-\frac{1}{2}$.
It means that, in accordance with the Harris-Aharony criterion,
the model can exhibit self-averaging properties. 
We study numerically the relative variances $R_{M}$ and $R_{\chi}$
for the probability $M$ of a site  belonging to the  "infinite" (maximum) 
cluster 
and the mean finite cluster size $\chi$. 
It was shown, that 
two-dimensional site-bound percolation on the square lattice, where the
bonds play the role
of impurity and the  sites play the role of the statistical ensemble,
over which the averaging is performed, exhibits self-averaging properties.

\end{abstract}
\maketitle


\newpage

\section{Introduction}
The influence of disorder on a phase transition
is one of the important problems in the theory of  phase transitions.  
In experimental measurements,  the thermodynamic
properties of one (or several) large-sized sample
(with respect to the number of molecules) are usually studied.
Therefore, it is important to know, whether a single large-sized sample 
 with quenched realization of impurities can 
represent the properties of a model?
The self-averaging properties of the system provide  
an  answer to this question.
If some quantity is self-averaging,
the measurement for a single large sample
gives a reasonable value for all samples of
such size. If the quantity is not self-averaging, 
an increase in  the system size does not 
make the measurement for a single sample representative.

Let us consider some statistical model with impurities
on the $d$-dimensional  lattice with a linear size $L$, in which we
average some quantity $X(\omega)$ for a
certain sample, where $\omega $ is the impurity realization 
from some impurity ensemble $\Omega$.
Here we assume that $X(\omega)$ is  the exact thermal average for impurity
realization $\omega$.
Let us denote by $[ X ] $ the average over all impurity realizations 
of ensemble 
$\Omega$:
$ [X]= \sum  \limits_{ \omega \in  \Omega } P( \omega )X( \omega )$,
where $P(\omega)$ is the probability of impurity realization
$\omega$.
Then, variance
$V_{X}=\sum \limits_{\omega \in \Omega} P(\omega)X^{2}(\omega) -[X]^{2}$.
Let us define the relative variance of $X$:
$R_{X}=\frac{V(x)}{[X]^{2}}$.
This quantity characterizes the size dependent properties of the model.
If $R_{X}(L) \sim L^{x},\;\;x=-d $ the quantity $X$
is said to be strongly self-averaging. If $x<0 $, $X$
said to be weak self-averaging, and if $\lim \limits_{L \to \infty}
R_{X}(L) \to {\rm const}$, than $X$ is non-selfaveraging.
We will now discuss 
 self-averaging of a disordered system; therefore
 we examine the averaging over all impurity
realizations, instead of
self-averaging of a pure system,
when we study the averaging over the whole 
statistical ensemble (for example, over all possible
spin configuration for a spin models).

The Harris criterion states~\cite{Harris} 
that the weak randomness does not change the critical behavior
of the $d$-dimensional second-order phase-transition model if
 the specific heat index $\alpha<0$, which corresponds to
the pure system correlation length index $\nu >2/d$. 
 As was first mentioned by Brout~\cite{Brout},
far for critiality where the
system size is much greater than 
the correlation length $L \gg \xi$, 
all additive thermodynamical quantities are
strongly self-averaging.
The self-averaging properties of a disordered system near the critiality
obeys  Harris-Aharony (HA) criterion see~\cite{HA1} and~\cite{HA2}.
This criterion  states that if randomness is irrelevant 
the system is governed by a pure fixed point and the relative variance
$R_{X} \sim L^{\frac{\alpha_{p}}{\nu_{p}}}$,
where $\alpha_{p}$ and $\nu_{p}$ are the critical exponents of a pure
system. If however, the system is governed by a random
fixed point, then $\lim \limits_{L \to \infty }R_{X}={\rm const}$.
This criterion explains the numerical results of many 
works~\cite{WD1,WD2,WD3}.   Wiseman  and  Domany in~\cite{WD3}
show that the  Ashkin-Teller model with $\alpha <0$
is weakly self-averaging. 

Two-dimensional percolation,
which can be treated as a $q$-state Potts model with $q \to 1$, 
is one of the most popular two-dimensional statistical models
with the second-order phase transition.
The two-dimensional percolation critical index $\alpha = -\frac{1}{2} <0$,
and, in accordance with the HA criterion we can expect that this
model exhibits weak self-averaging properties.
We study numerically the relative variances $R_{M}$ and $R_{\chi}$
of the probability $M$ of a site to belong the  maximum cluster  (the 
analog of magnetization)
and the mean finite   cluster size $\chi$ (the analog of the magnetic
susceptibility). 
It was shown, that 
two-dimensional site-bound percolation, where  bonds play the role
of impurities and sites play the role of statistical ensemble
over which the averaging is performed, exhibits self-averaging
properties.
The article is arranged as follows.
In  Section~2 we describe the site-bond percolation model and 
two types of impurities distribution;
in  Section~3 we discuss the self-averaging
criterion and its phenomenological derivation; 
in Section~4 we present the numerical results
and Section~5 is the conclusion.

\section{Two-dimensional site-bond percolation model and canonical and 
grand-canonical impurity
distributions}

Site-bond percolation is a combination of purely site percolation and
purely bond percolation. In purely site percolation, one investigates
clusters of occupied sites. In purely bond percolation,
one investigates clusters of sites, connected by occupied bonds.
In site-bond percolation one, investigates clusters of 
occupied sites connected by occupied bonds.
Each site is occupied with probability $p_{s}$, and
each bond is occupied with probability $p_{b}$.
In critical point $(p_{s}^{*},p_{b}^{*})$
the correlation length becomes infinite.
Yanuka and Engelman~\cite{YE} proposed approximate formula
for a critical curve in $(p_{s},p_{b})$ plane
$$
\frac{\log(p_{s})}{\log(p_{s}^{*})}+
\frac{\log(p_{b})}{\log(p_{b}^{*})}=1
$$
The two-dimensional site-bond percolation
belongs the same universality class as the
two-dimensional purely site percolation.
Let us define by $n_{C}$ the mean number of clusters of size $C$
per lattice site. Then, the probability of a site to belong to the maximum 
cluster 
is 
\begin{equation}
\label{wqm}
M=n_{C_{max}}C_{max} 
\end{equation}
and the mean finite cluster size 
\begin{equation}
\label{wqx}
\chi=\frac{ \sum \limits_{C,C\ne C_{max}} n_{C}C^{2} }{\sum 
\limits_{C, 
C\ne C_{max}} 
n_{C} C}
\end{equation}
Near the critical point $(p_{s}^{*},p_{b}^{*})$
this quantities scales as follows
\begin{equation}
\label{eqmscal}
M(p_{s},p_{b}=p_{b}^{*}) \sim (p_{s}-p_{s}^{*})^{\beta}, \;\;p_{s} >
p_{s}^{*},\;\; \beta=\frac{1}{18}
\end{equation}
and
\begin{equation}
\label{eqxscal}
\chi(p_{s},p_{b}=p_{b}^{*}) \sim (p_{s}-p_{s}^{*})^{-\gamma},
\;\;\gamma=\frac{43}{18}
\end{equation}
The correlation length $\xi(p_{s},p_{b}=p_{b}^{*}) \sim 
(p_{s}-p_{s}^{*})^{-\nu}$ scaling index 
is $\nu=\frac{4}{3}$.

As mentioned above, each bond is occupied with
probability $p_{b}$ and is empty with probability $1-p_{b}$;
therefore the total number of occupied bonds on the lattice
fluctuates. This method of generating
 impurity configurations
is known as Grand Canonical (GC) by analogy with Grand Canonical 
statistical ensemble with a fluctuating number of particles.
However we can fix the number of occupied bonds $N_{b}$
and then distribute them randomly on the lattice.
This method of generating   
 impurity configurations
is known as  Canonical (C) by analogy with the Canonical
statistical ensemble with fixing number of particles.

\section{Self-Averaging criterion}
\label{secsa}

Let us explain the HA criterion via simple phenomenological
considerations~\cite{WD2}. In this section, we use the $T$  temperature 
as a parameter of the model  
but keeping in mind, that in the case of percolation,
the role of this parameter is played by the site concentration $p_{s}$. 

We characterize every sample (impurity realization) 
$\omega$ with size $L$ by a pseudo-critical temperature 
$T^{*}(\omega,L)$. $T^{*}(\omega,L)$ fluctuates about its mean
value and is averaged over all impurity configurations
 $T^{*}=[T^{*}(\omega,L)]$. 
We introduce the reduced 
temperature for each sample
\begin{equation}
\label{eqrt}
\dot t_{\omega}=\frac{T-T^{*}(\omega,L)}{T^{*}}
\end{equation}
In the vicinity of the critical point, 
the quantity $X$
scales as
\begin{equation}
\label{eqxsc}
X_{\omega}(T,L)=L^{\rho} \tilde Q_{\omega}(\dot t_{\omega} L^{y_{t}})
\end{equation}
Here, $\rho$ is the exponent characterizing the behavior
of $[X]$ at $T^{*}$ and the thermal scaling index  
$y_{t}=\frac{1}{\nu}$ is assumed to be universal for all samples. 
The form of function $\tilde Q_{\omega}$ is assumed to be sample 
dependent. We assume that (this relation is numerically checked in the 
next section) 
\begin{equation}
\label{eqtc}
\left( \delta T^{*} \right)^{2}\sim \left( \delta \dot t_{\omega} 
\right)^{2} \sim 
L^{-d}
\end{equation}
So the variance of the argument of the function $\tilde Q_{\omega}$
scales as 
\begin{equation}
\label{eqarg}
\left( \delta \dot t_{\omega} L^{y_{t}}\right)^{2} \sim   
L^{-d+\frac{2}{\nu}}
\end{equation}
We keep in mind that the scaling relation~\cite{DM} $\alpha=2- \nu d$.   
So, the variance $V_{x}$ of quantity $X$ at the critical point 
scales as 
\begin{equation}
\label{eqvar}
V_{X} \sim L^{2 \rho}\left( \delta \dot t_{\omega} L^{y_{t}}\right)^{2} 
  \sim
L^{2 \rho-d+\frac{2}{\nu}}=L^{2 \rho+\frac{2-d \nu}{\nu}}=L^{2 
\rho+\frac{\alpha}{\nu}}
\end{equation}
and the relative variance ($[X]\sim L^{\rho}$) is
\begin{equation}
\label{eqrelvar}
R_{X}=\frac{V_{X}}{[X]^{2}}=L^{2  \rho+\frac{\alpha}{\nu}-2 \rho}=
     L^{\frac{\alpha}{\nu}}
\end{equation}
 We thus obtain the Harris-Aharony criterion:
{\it if the scaling index $\alpha =const<0$, then the model is 
self-averaging
and  the relative variance 
 for quantity $X$ is
$R_{X}\sim 
L^{\frac{\alpha}{\nu}}$} 
from simple scaling relations.

\section{Numerical results}

In our computation, the bond play the role of impurities;
therefore, we fix the bond concentration $p^{*}_{b}=0.875$ and 
consider the site concentration $p_{s}$ as parameter
of our model.

\begin{enumerate}
\item we generate (by  C and  GC methods)
the bond configuration $\omega$.
\item for this bond configuration 
we  generate a set of site configurations (for site we use only GC 
method);
For each site configuration we, calculate the probability of a site
 belonging to the maximum 
cluster  and
the mean finite cluster size;
\item We average these quantities over site realizations
and obtain mean values $M(p_{s},p_{b})$ and $\chi(p_{s},p_{b})$.
We split the set of site configurations into ten series
to evaluate the numerical inaccuracy $\Delta M$ and $\Delta \chi$ 
\item we perform steps 1.--3. for another  bond configuration.
\end{enumerate}

The $\chi_{\omega}(p_{s})$ dependence  for three 
different 
bond realizations $(\omega=1,2,3)$ is shown in Fig.\ref{figchi}.
Let us calculate the pseudocritical site concentration $p^{*}_{s}(\omega)$
for each bond realization $\omega$. We assume, that
 the mean finite cluster size
$\chi_{\omega}(p_{s})$ has maximum at  the
pseudocritical point $p^{*}_{s}(\omega)$,
and therefore, we approximate the data for $\chi_{\omega}$
 near the maximum of $\chi_{\omega}$ by the parabola $\chi_{\omega} 
(p_{s}) \simeq a- b 
\left(p_{s}-p^{*}_{s}(\omega) \right)^{2}$
and treat $p^{*}_{s}(\omega)$ as the pseudocritical concentration for
bond realization $\omega$. The locations of pseudocritical points
for bond realizations $\omega=1,2,3$ is shown in Fig.\ref{figchi}
by vertical lines.

Let us find the critical point $p^{*}_{s}$. We calculate 
 the mean critical site concentration $[p^{*}_{s}(L)]$
averaged over $100$ bond realizations
created by GC and C methods as a function of lattice size $L$.
These numerical data and results of approximation 
\begin{equation}
\label{eqdatpc}
\begin{array}{rl}
{\rm C-method}: & [p^{*}_{s}(L)]_{C}\simeq 
0.6519(5)-0.34(10)L^{-0.94(9)}\\
{\rm GC-method}:& [p^{*}_{s}(L)]_{GC}\simeq 
0.6511(4)-0.39(15)L^{-1.02(11)}\\
\end{array}
\end{equation} 
is shown in Fig.\ref{figpc}.
We take the concentration $p^{*}_{s}=0.6515$ (averaged by C and GC 
methods) as critical. 
The variance of pseudocritical concentration behaves
as $(\delta p_{s}^{*})^{2} \sim L^{-2}$ (Fig.\ref{figdt}),
as we assumed in the previous section. 

Let us investigate the self-averaging properties of $M$ and $\chi$
at the critical  point $(p^{*}_{b}=0.875, p^{*}_{s}=0.6515)$.

Now we describe computational procedure that we use to calculate 
the relative variance. Here, we follow the method 
described~\cite{WD1,WD2}.
First, we note that, for each impurity realization 
$\omega$, instead of the exact value of quantity $X_{\omega}$
we get some value $ \bar X_{\omega}$
averaged over site configurations
with a  numerical error  
\begin{equation}
\left( \delta \bar X_{\omega} 
\right)^{2}=\frac{\sigma_{p_{b},\omega}}{N_{s}/\tau_{\omega}}
\end{equation}
Here, $N_{s}$ is the number of site configurations (the length
of Monte-Carlo run) and $\tau_{\omega}$ is autocorrelation time.
To calculate $\left( \delta \bar X_{\omega} \right)^{2}$
we split the MC sequence of site configurations of lenth $N_{s}$
into 10 subsequences and treat each subsequence as independent.
We define by $[\dots ]$ the  averaging over impurity -- bond 
configurations.
Thus, the error of $[\bar X_{\omega}]$ averaged over $N_{b}$
bond configurations is
\begin{equation}
\left( \delta [\bar X_{\omega}]\right)^{2}= 
\frac{1}{N_{b}-1} \sum \limits_{\omega \in \Omega}
\left( [\bar X_{\omega}^{2}]-[\bar X_{\omega}]^{2} \right)
\end{equation}
The total total error $\left( \delta [\bar X_{\omega}]\right)^{2}$
has two contributing terms: from the sample to sample fluctuation
of exact $X_{\omega}$ about $[X_{\omega}]$ and
from the fluctuations of $\bar X_{\omega}$, averaged over
 the finite number of spin configurations, about $X_{\omega}$
for each bond realization $\omega$.  
\begin{equation}
\left( \delta [\bar X_{\omega}]\right)^{2}=
\frac{V_{X}}{N_{b}}+ \left[ 
\frac{\sigma_{p_{b},\omega}}{N_{b}N_{s}/\tau_{\omega}}\right]
\end{equation}
Here, $N_{b}$ is the number of bond configurations.
We can calculate 
$\left[ \frac{\sigma_{p_{b},\omega}}{N_{b}N_{s}/\tau_{\omega}}\right]=
\frac{1}{N_{b}} \left[ \left( \delta \bar X_{\omega}  \right)^{2}  
\right]$
by averaging the error $\left(\delta \bar X_{\omega}  \right)^{2}$ over 
the bond configurations $\omega \in \Omega$.
So, we can express the relative variance via 
$\left( \delta [\bar X_{\omega}]\right)^{2} $ and 
$\left[(\delta \bar X_{\omega} )^{2} \right]$ 
\begin{equation}
\label{eqr}
R_{X}=\frac{V_{X}}{[ \bar X_{\omega}]^{2}}=
\frac{1} { [\bar X_{\omega} ]^{2} }
\left( N_{b} \left( \delta [\bar X_{\omega}] \right)^{2}-
\left[(\delta \bar X_{\omega} )^{2} \right]
\right)
\end{equation}
The numerical data for relative variance $R_{M}$ and 
$R_{\chi}$ computed by the
  C and GC methods  in accordance~(\ref{eqr}),
 are shown in Fig.\ref{figrd}
and~\ref{figrdm} respectively. We can see that 
points lies on the straight lines in log-log
scale. Thus, we approximate the data by power function $a L^{b}$
on the interval $L \in [48,128]$. The results of approximation
are plotted in  Fig.\ref{figrd}
and~\ref{figrdm} and placed below. As we might expect,
for the GC method the relative variance is greater, than for the C-method
because of the fluctuation of bonds number.
$${\rm C}: R_{M}\simeq 0.0041(3)L^{-0.504(19)}$$
$${\rm GC}: R_{M}\simeq 0.0147(15)L^{-0.50(2)}$$
$${\rm C}: R_{\chi}\simeq 0.0243(2)L^{-0.53(2)}$$
$${\rm GC}: R_{\chi}\simeq 0.0259(5)L^{-0.51(5)}$$
We see, that in  excellent agreement with HA criterion, 
the relative variance of measured quantities obeys the power low 
dependence 
$R \sim L^{-\frac{1}{2}} =L^{\frac{\alpha}{\nu}}$.
As a result, we can state that this model is self-averaging.

\section{Conclusion}

We have shown numerically that the two-dimensional site-bond
percolation exhibit self-averaging properties at critical point. In our 
numerical experiments
 bonds play  the role of the quenched disorder, 
and the sites play role of  the statistical ensemble.
We have found find that relative variance 
scales $R_{M} \sim L^{-\frac{1}{2}}$, $R_{\chi} \sim L^{\frac{1}{2}}$,
where $-\frac{1}{2}=\frac{ \alpha}{\nu} $.
We assume that we can consider the sites as a quenched
disorder and bonds, as a statistical ensemble,
and the results will be the same.

We can also expect for the site percolation that, if we "freeze" the
state of some sites  and
average over other sites, we get the same weakly self-averaging
behavior with respect this "frozen" sites, which
play the role of quenched disorder.
Of cause, the same is valid for bond percolation.

I am  grateful  to the Joint SuperComputer Center RAN (www.jscc.ru) for
providing computational resources.


\clearpage

\centerline{\Large\bf Figure captions}
\bigskip
\bigskip

\begin{fig}
\label{figchi}
The mean finite-cluster size $\chi(p)$
for three realization of impurity configurations. The critical
concentrations $p^{*}_{s}$ for each realization $\omega$, 
 are shown by vertical lines.
\end{fig}

\bigskip

\begin{fig}
\label{figpc}
The mean pseudocritical site concentration $[p^{*}_{s}]$
as a function of lattice size $L$
for  bond realizations, created by C and GC methods,
the results of approximation by a power function of $L$,
and the extrapolated critical concentration $p_{s}^{*}=0.6515$
\end{fig}

\bigskip

\begin{fig}
\label{figdt}
The variance $\left(\delta p^{*}_{s} \right)^{2}$ of critical 
concentration 
as a function of lattice size $L$ for C 
(crosses) and GC (triangles) distributions, and 
the results of approximation by function $a L^{-b}$.
\end{fig}

\bigskip

\begin{fig}
\label{figrdm}
The relative variance $R_{M}$ of probability $M$ of a site belonging to 
the maximum cluster 
as a function of lattice size $L$ for GC 
(crosses) and C (triangles) distributions and 
the results of approximation by function $a L^{-b}$.
\end{fig}

\bigskip

\begin{fig}
\label{figrd}
The relative variance $R_{\chi}$ of the mean finite-cluster size $\chi$ 
as a function of lattice size $L$ C  
(crosses) and GC (triangles) distributions and 
the results of approximation by function $a L^{-b}$.
\end{fig}

\begin{figure}[p]
\vspace{50mm}
\begin{center}
\epsfxsize=150mm
\epsfysize=100mm
\begin{picture}(0,-00)
\setlength{\unitlength}{1.000pt}
\put(295,253){ $\omega=1$}
\put(295,236){ $\omega=2$}
\put(295,220){ $\omega=3$}
\put(230,-10){\Large $p_{s}$}
\put(-30,168){\LARGE $\chi(p_{s})$}
\end{picture}%
\epsfbox{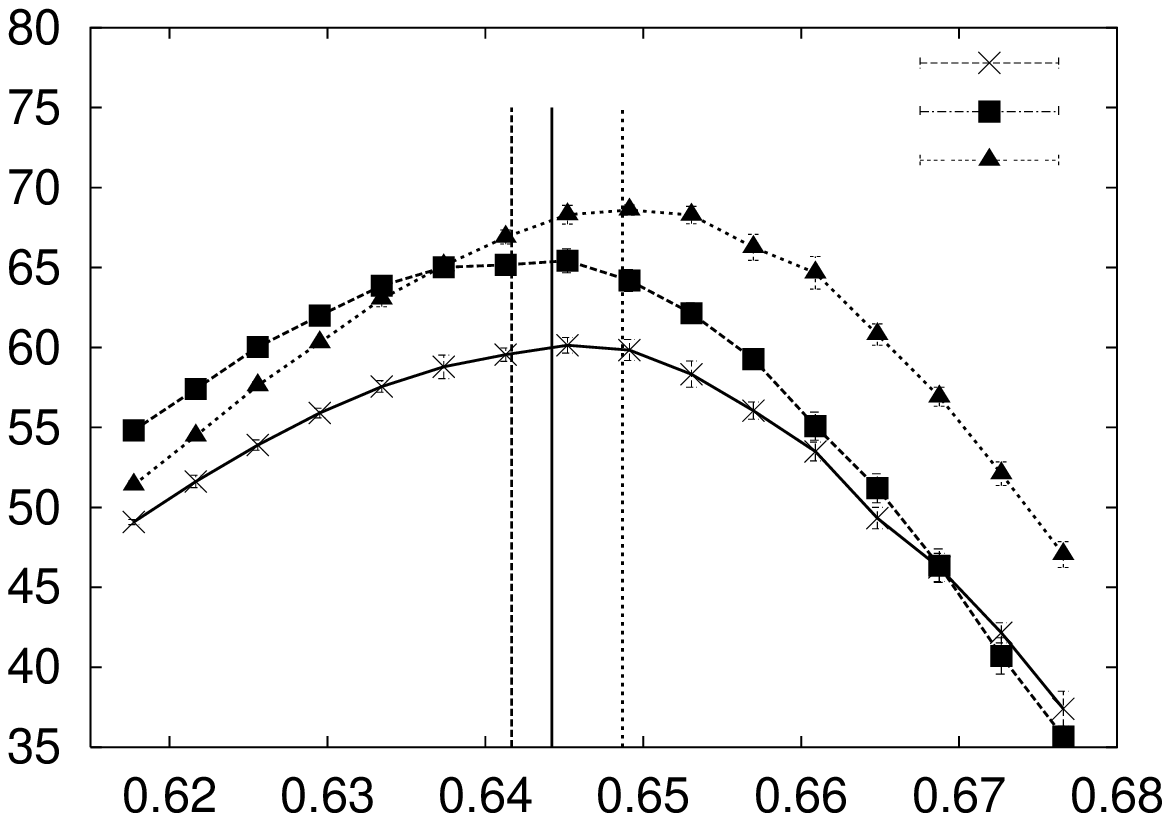}
\end{center}
\vspace{0mm}
\caption{ }
\end{figure}

\begin{figure}[p]
\vspace{50mm}
\begin{center}
\epsfxsize=150mm
\epsfysize=100mm
\begin{picture}(0,-00)
\setlength{\unitlength}{1.000pt}
\put(255,105){C-method $[p^{*}_{s}]$}
\put(242,88){ GC-method $[p^{*}_{s}]$}
\put(178,71){ $0.6519(5)-0.34(10)L^{-0.94(9)}$}
\put(175,55){ $0.6511(4)-0.39(15)L^{-1.02(11)}$}
\put(260,40){ $p^{*}_{s}=0.6515$}
\put(230,-10){\Large $L$}
\put(-30,178){\LARGE $[p^{*}_{s}](L)$}
\end{picture}%
\epsfbox{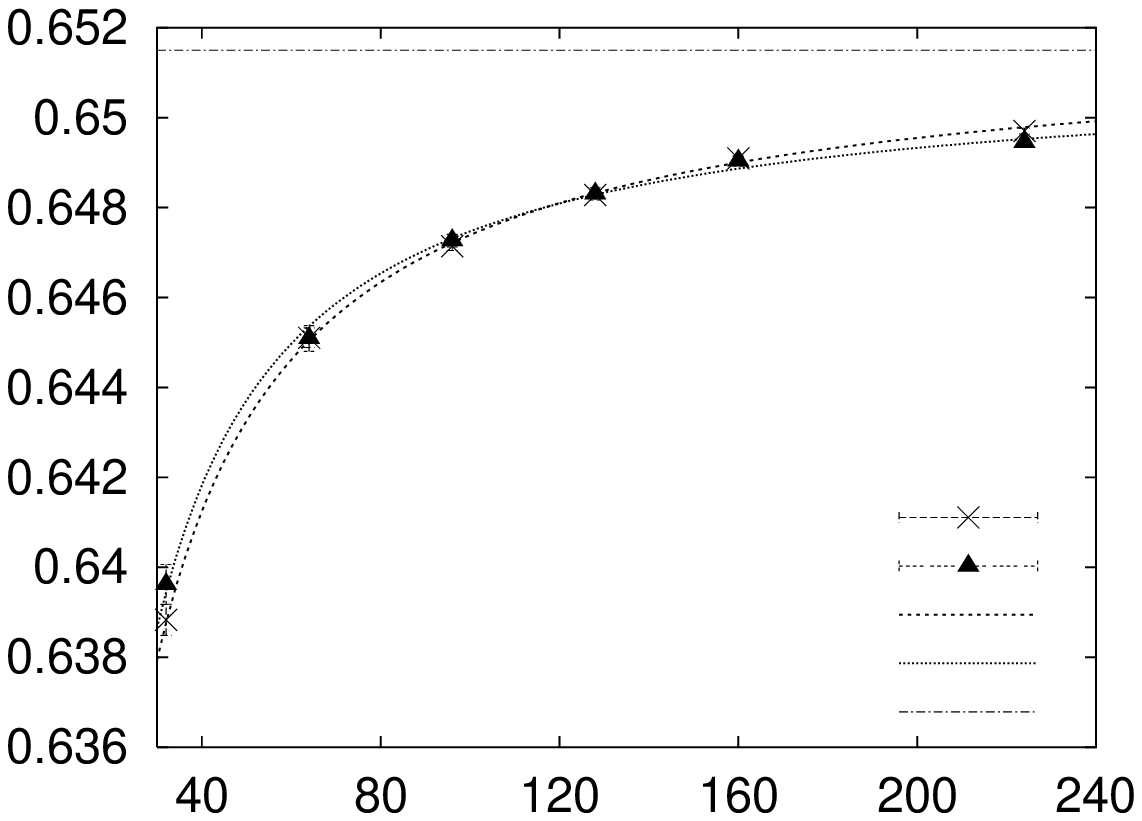}
\end{center}
\vspace{0mm}
\caption{ }
\end{figure}

\begin{figure}[p]
\vspace{50mm}
\begin{center}
\epsfxsize=150mm
\epsfysize=100mm
\begin{picture}(0,-00)
\setlength{\unitlength}{1.000pt}
\put(244,253){C-method $\left(\delta p^{*}_{s} \right)^{2}$}
\put(234,238){GC-mathod $\left(\delta p^{*}_{s} \right)^{2}$}
\put(241,222){ $0.0014(7)L^{-2.1(1)} $}
\put(238,205){ $0.0020(5)L^{-1.93(6)}$}
\put(230,-10){\Large $L$}
\put(25,168){\Large $\left(\delta p^{*}_{s} \right)^{2}$}
\end{picture}%
\epsfbox{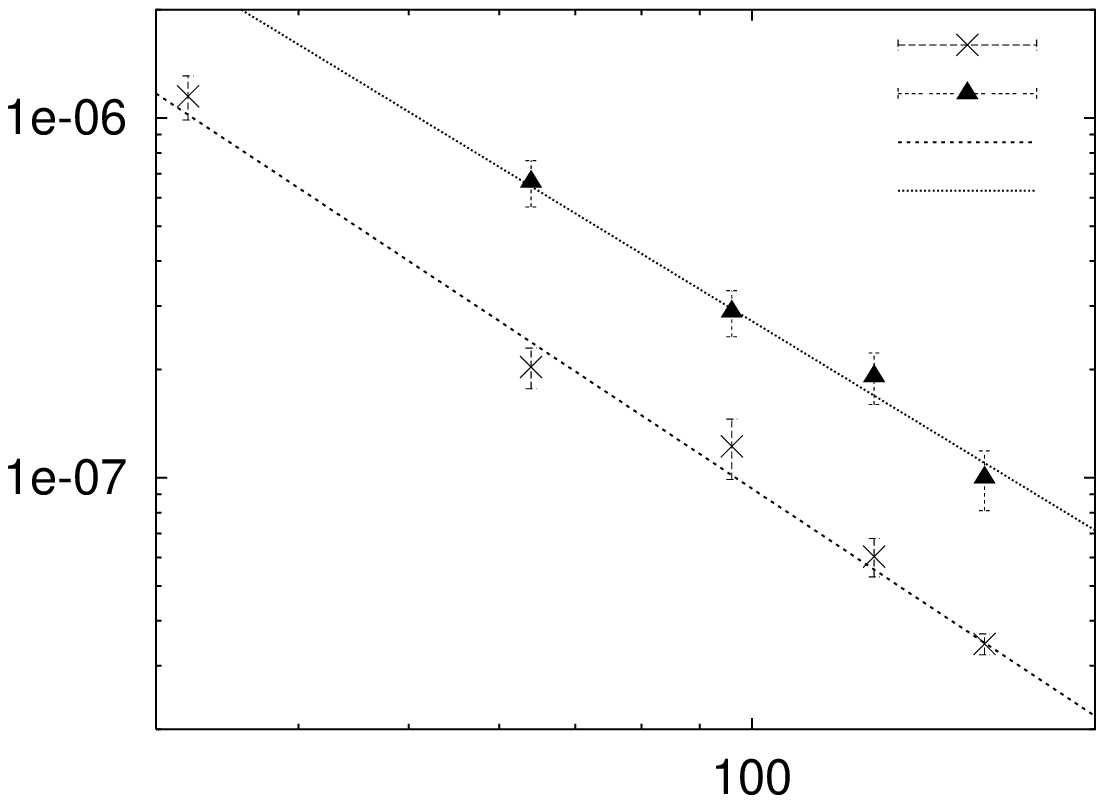}
\end{center}
\vspace{0mm}
\caption{ }
\end{figure}

\begin{figure}[p]
\vspace{50mm}
\begin{center}
\epsfxsize=150mm
\epsfysize=100mm
\begin{picture}(0,-00)
\setlength{\unitlength}{1.000pt}
\put(252,253){C-method $R_{M}$}
\put(243,239){GC-mathod $R_{M}$}
\put(229,219){ $0.0041(3)L^{-0.504(19)} $}
\put(231,201){ $0.0147(15)L^{-0.50(2)}$}
\put(230,-10){\Large $L$}
\put(25,168){\Large $R_{M}$}
\end{picture}%
\epsfbox{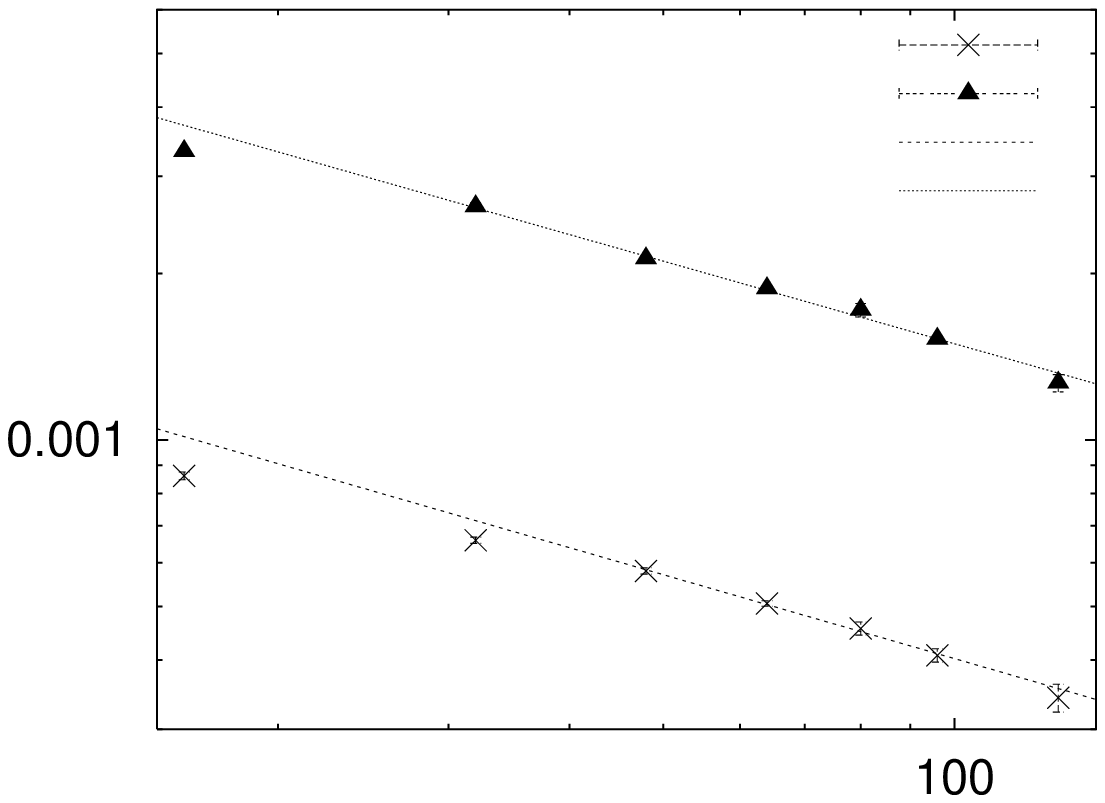}
\end{center}
\vspace{0mm}
\caption{ }
\end{figure}

\begin{figure}[p]
\vspace{50mm}
\begin{center}
\epsfxsize=150mm
\epsfysize=100mm
\begin{picture}(0,-00)
\setlength{\unitlength}{1.000pt}
\put(253,251){C-method $R_{\chi}$}
\put(244,236){GC-method $R_{\chi}$}
\put(237,219){ $0.0243(2)L^{-0.53(2)} $}
\put(237,203){ $0.0259(5)L^{-0.51(5)}$}
\put(230,-10){\Large $L$}
\put(25,148){\Large $R_{\chi}$}
\end{picture}%
\epsfbox{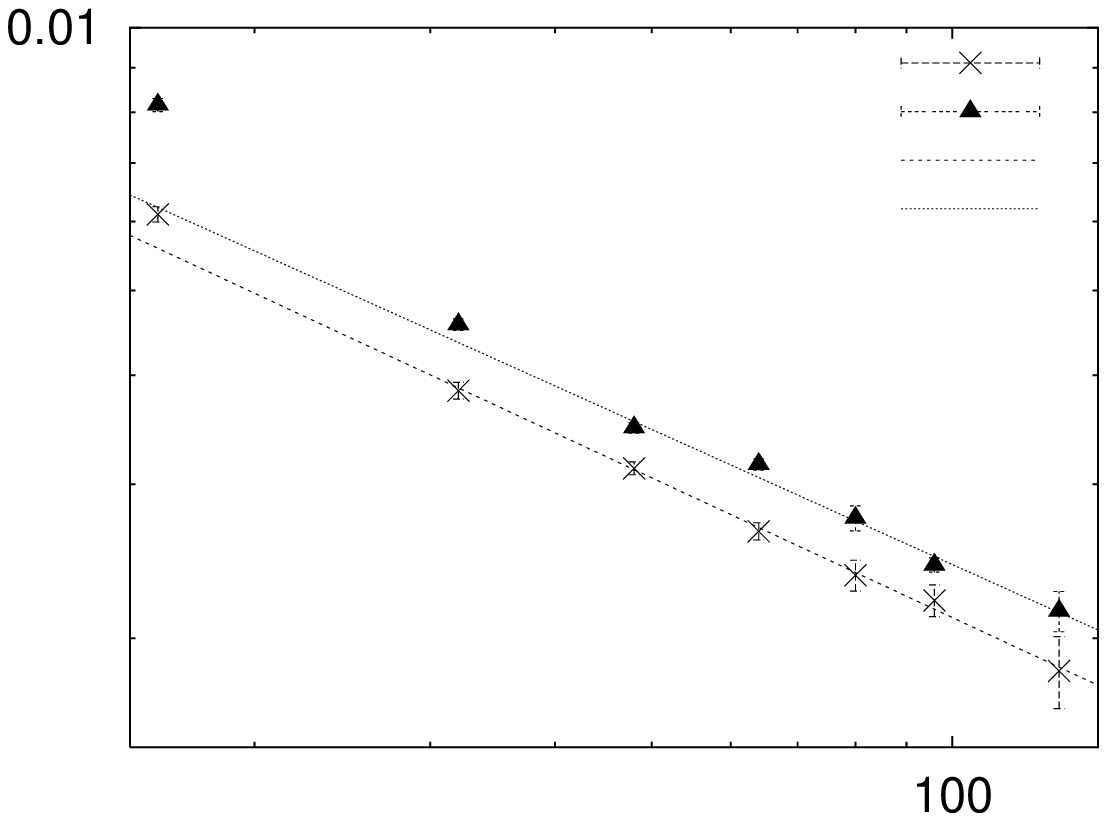}
\end{center}
\vspace{0mm}
\caption{ }
\end{figure}

\end{document}